\documentclass{moriond}

\def\be{\begin{equation}}
\def\ee{\end{equation}}
\def\bea{\begin{eqnarray}}
\def\eea{\end{eqnarray}}

\newcommand{\Photo}{}

\usepackage{xspace}
\usepackage{amsmath}
\newcommand{\AtlasMC}[2][]{%
  \if\relax\detokenize{#1}\relax
    \textsc{#2}\xspace%
  \else
    \textsc{#2}\,#1\xspace%
  \fi
}

\newcommand{\MADGRAPH}[1][]{\AtlasMC[#1]{MadGraph}}
\newcommand{\MADSPIN}{\AtlasMC{MadSpin}}
\newcommand{\HATHOR}{\AtlasMC{Hathor}}
\newcommand{\MGNLO}[1][]{\AtlasMC[#1]{MadGraph5\_aMC@NLO}}

\newcommand{\Pythia}[1][]{\AtlasMC[#1]{Pythia}}
\newcommand{\Herwig}[1][]{\AtlasMC[#1]{Herwig}}

\newcommand{\POWHEG}[1][]{\AtlasMC[#1]{Powheg}}

\newcommand{\MiNNLO}{\texttt{MiNNLOps}\xspace}
\newcommand{\hvq}{\texttt{hvq}\xspace}
\newcommand{\ttbar}{\ensuremath{t\bar{t}}\xspace}

\newcommand{\bbfourl}{\ensuremath{bb4\ell}\xspace}
\newcommand{\mttbar}{\ensuremath{m_{\ttbar}}\xspace}
\newcommand{\TeV}{\ensuremath{\text{Te\kern -0.1em V}}}
\newcommand{\GeV}{\ensuremath{\text{Ge\kern -0.1em V}}}
\newcommand{\ifb}{\mbox{fb\(^{-1}\)}}
\newcommand{\ipb}{\mbox{pb\(^{-1}\)}}

\newcommand{\pT}{\ensuremath{p_\mathrm{T}}\xspace}

\begin{document}
\vspace*{4cm}
\title{ATLAS and CMS measurements of the $\ttbar$ cross section,\\ including off-shell and near threshold}

\author{Baptiste Ravina, on behalf of the ATLAS and CMS Collaborations}

\address{CERN, CH-1211 Geneva, Switzerland}

\maketitle\abstracts{
Recent measurements of the \ttbar cross section, performed both inclusively and differentially by the ATLAS and CMS Collaborations, are reported.
In particular, off-shell effects are probed in the $pp\to W^+bW^-\bar{b}$ and $pp\to e^\pm\mu^\mp +b\bar{b}$ processes, and modelling aspects of the \POWHEG \bbfourl Monte Carlo generator are discussed.
Cross section and properties measurements are also performed at the threshold: we review an indirect extraction of the top quark Yukawa coupling, as well as the recent observations by both experiments of an excess of events near the top pair production threshold that is consistent with the formation of quasi-bound states.
}

\section{Introduction}

The Large Hadron Collider (LHC) at CERN produced more than 15 top-antitop-quark (\ttbar) events per second during its Run 2, making it a veritable ``top quark factory''.
The impressively large datasets collected by both the ATLAS and CMS experiments during this period, 2015 to 2018, amount to more than 115 million pairs of top quarks.
This has in turn enabled precise measurements of the \ttbar cross section, reported in Section~\ref{sec:xsec}, using either dedicated low-pileup environments~\cite{CMS:2024ghc} or the very well-controlled dileptonic final state~\cite{ATLAS:2025iza}.

Beyond measurements of the doubly-resonant final state, subtleties of the off-shell topologies, such as $W^+bW^-\bar{b}$ or $e^\pm\mu^\mp +b\bar{b}$, are revealed in dedicated analyses~\cite{ATLAS:2025iza,ATLAS:2025fhs}.
As shown in Section~\ref{sec:off}, they highlight in particular the need for precise modelling of these processes in Monte Carlo (MC) generators~\cite{ATL-PHYS-PUB-2025-036}.

Near the top quark pair production threshold, $\mttbar\sim 2m_t$, fundamental aspects of quantum chromodynamics (QCD) and the Standard Model can be probed.
Top quarks moving at low velocities relative to each other are able to exchange soft gluons and virtual Higgs bosons, imprinting the large top Yukawa coupling ($g_t$) and the effects of extremely short-lived quasi-bound states onto key distributions of their decay products.
Tailored efforts to uncover these effects~\cite{ATLAS:2025ciy,CMS:2025kzt,ATLAS:2026dbe} are presented in Section~\ref{sec:thresh}.

\section{Precise measurements of the \ttbar cross section}
\label{sec:xsec}

A first approach, followed by the CMS Collaboration~\cite{CMS:2024ghc}, uses a small but low-pileup dataset of \ttbar events.
These proton-proton collisions, collected in 2017 at a centre-of-mass energy $\sqrt{s}=5.02~\TeV$, feature a remarkably low average number of interactions per bunch crossing, $\langle\mu\rangle =2$.
By comparison, $\langle\mu\rangle =33$ with the CMS detector under standard data-taking conditions in Run 2.
This makes for a clean environment, where the hard-scatter process can be studied in almost complete isolation from background pileup contamination.
On the other hand, only 302~\ipb\ of data were recorded under these conditions.

In this approach, the semi-leptonic \ttbar final state is targeted, to benefit from the increased branching ratio compared to a previous measurement in the dileptonic $e\mu+$jets channel.
Background contributions include $W+$jets production, which is reduced using multivariate techniques, as well as the singly-resonant $tW$ production mode (obtained from MC simulations) and a minor QCD multi-jet background, estimated with data-driven techniques.
A maximum likelihood fit is performed in bins of jet and $b$-jet multiplicities, yielding an inclusive cross section
\begin{equation}
       \sigma(\ttbar) = 62.5\pm 1.6~\text{(stat.)}~^{+2.6}_{-2.5}~\text{(syst.)}~\pm 1.2~\text{(lumi.)}~\text{pb},
\end{equation}
in reasonable agreement with the theoretical prediction $\sigma(\ttbar)=69.5^{+3.5}_{-3.7}~\text{pb}$ at next-to-next-to-leading order (NNLO) in QCD and next-to-next-to-leading logarithmic (NNLL) accuracy.
The leading systematic uncertainties on this result arise from the matrix-element to parton-shower matching component, as well as from the $b$-tagging and trigger efficiencies, owing to the limited dataset.

An alternative approach is employed by the ATLAS Collaboration~\cite{ATLAS:2025iza}, focusing instead on the low-branching-ratio but extremely clean final state offered by an $e\mu+b\bar{b}$ event selection.
The lower branching ratio is compensated by using the full Run 2 dataset, with an integrated luminosity of $140~\ifb$, yielding about a million \ttbar events after selection.
Previous results from the ATLAS Collaboration on this same dataset have introduced the various ingredients needed for a precision measurement: a $b$-jet counting technique, which allows for an in-situ extraction of $b$-tagging efficiencies, as well as dedicated leptonic isolation efficiencies obtained in the busier \ttbar environment.
Additionally, the \POWHEG \MiNNLO event generator is employed to benefit from NNLO-accurate modelling of the leptons' \pT and to improve the acceptance.
An inclusive cross section
\begin{equation}
       \sigma(\ttbar) = 829.3\pm 1.3~\text{(stat.)}~\pm 8.0~\text{(syst.)}~\pm 7.3~\text{(lumi.)}~\pm 1.9~\text{(beam)}~\text{pb}
\end{equation}
is obtained, in excellent agreement with the NNLO+NNLL prediction of $\sigma(\ttbar)=834^{+29}_{-37}~\text{pb}$.

The aforementioned $b$-jet counting technique is promoted to run over bins of various leptonic observables, allowing for one-dimensional and two-dimensional measurements of the (normalised) differential \ttbar cross sections.
These unfolded cross sections are limited in part by the statistical uncertainty of the data, but also by the interference between the \ttbar and $tW$ processes at high-\pT, and by various aspects of \ttbar MC modelling at low-\pT.
The description of the data by several MC setups is quantified as $\chi^2/$NDF tables, and reveal some consistent patterns across the unfolded observables:
\begin{itemize}
       \item the baseline NLO-accurate \POWHEG \hvq setup, whether showered by \Pythia[8] or \Herwig[7], shows poor modelling of observables sensitive to the \pT of the top quark;
       \item this mis-modelling is corrected by applying kinematic reweighting of \hvq to NNLO QCD + NLO EW predictions (using the parton-level top and anti-top \pT distributions);
       \item an alternative NLO-accurate setup generated with \MGNLO+\Pythia[8] (without higher-order corrections) performs similarly to or better than \hvq;
       \item the NNLO-accurate \POWHEG \MiNNLO + \Pythia[8] setup (without NLO EW corrections) generally performs best across all 1D and 2D observables;
       \item the doubly-differential distribution of $\Delta\phi^{e\mu}$ versus $m^{e\mu}$ remains difficult to model, even for \MiNNLO with a $\chi^2/$NDF$=50.2/39$.
\end{itemize}

Related to that last point, it is worthwhile to note that the mismodelling in the $\Delta\phi^{e\mu}\times m^{e\mu}$ distribution is driven in part by the normalisation of the first bin of $m^{e\mu}$, corresponding to $m^{e\mu}<80~\GeV$.
This selection reflects the \ttbar threshold, and it is observed that the data is in excess of all MC generators tested in this analysis.

\section{Challenges in off-shell modelling}
\label{sec:off}

\subsection{Resonant and non-resonant production processes}

A limitation for precise \ttbar differential cross section measurements arises from the treatment of off-shell backgrounds and interference effects.
For instance, Born-level contributions to the $\ell^\pm\bar{\nu}b\ell^{'\mp}\nu\bar{b}$ final state include diagrams with two resonant top quarks (\ttbar), a single resonant top quark and a $Wb$ pair ($tW$), or no top quark resonance at all ($WbWb$).
In the experimental analyses discussed so far, these three types of diagrams are modelled using dedicated MC simulations.
However, at NLO in QCD, the $tWb$ process overlaps and interferes with \ttbar production.
This effect is modelled using the so-called ``Diagram Removal'' (DR) and ``Diagram Subtraction'' (DS) schemes, which are non-gauge-invariant procedures and can lead to large uncertainties in particular regions of phase-space (e.g. comparing the predictions of $\ttbar + tW$ DR to $\ttbar +tW$ DS).
Furthermore, both the \ttbar and $tW$ simulations start from a matrix-element generation that involves on-shell top quarks, which are subsequently decayed to final state products.
This decay relies on the narrow-width approximation (NWA), of order $\mathcal{O}(\Gamma/M)=0.8\%$ for the top quark and $2.6\%$ for the $W$ boson, which is implemented in both the \POWHEG \hvq and the \MADSPIN programmes; the difference between these two implementations is sometimes considered an additional ``lineshape'' modelling uncertainty.

The \POWHEG authors introduced an alternative approach (``\bbfourl'') to the modelling of all these effects, starting from the $2\to 6$ matrix element $pp\to \ell^\pm\bar{\nu}b\ell^{'\mp}\nu\bar{b}$ and implementing NLO QCD corrections in a MC simulation matched to parton showers.
In this way, the full interference between $tW$ and \ttbar production is taken into account, as well as off-shell effects in the decays, subleading electroweak production diagrams, and the treatment of spin correlations is NLO-accurate.
From an experimental point of view, this \POWHEG \bbfourl MC simulation removes the need for the ad-hoc ``lineshape'' and ``DR/DS'' modelling uncertainties.

As discussed in a recent ATLAS public note~\cite{ATL-PHYS-PUB-2025-036}, the combination of these systematic uncertainties can have effects up to $10\%$ in the tails of the top quark mass distribution.
However, while the benefit from more accurate matrix elements seems immediate, care is needed when devising prescriptions for modelling uncertainties in experimental settings.
In particular, parton shower matching, e.g. to \Pythia[8], behaves differently as the first gluon emission from the top quark decay can now already appear in the \bbfourl matrix element, whereas it was entirely modelled by the parton shower algorithm when interfaced to \hvq.
Turning to the $b$-jet fragmentation functions, it is observed that the effects of splitting kernel variations in \Pythia[8] are less pronounced for \bbfourl than for \hvq, but the former setup gains sensitivity to variations of $\alpha_S(m_Z)$ in the \POWHEG Sudakov corrections.
The ATLAS public note~\cite{ATL-PHYS-PUB-2025-036} therefore lays out an updated set of systematic prescriptions, which will be employed in future top quark measurements sensitive to an NLO-accurate description of the decay products.
It is worth noting that NNLO QCD corrections to the $pp\to \ell^\pm\bar{\nu}b\ell^{'\mp}\nu\bar{b}$ are currently unknown, but active progress is being made on the theory side, such that it is reasonable to expect that approximate NNLO QCD corrections could be available within the next couple of years, allowing for kinematic reweighting of the \bbfourl MC simulations.

\subsection{Experimental measurements of off-shell processes}

The $WbWb$ process has been studied in the semi-leptonic channel by the ATLAS Collaboration~\cite{ATLAS:2025fhs}, providing three sets of measurements unfolded to stable-particle level.
An ``inclusive'' analysis is performed, with the largest fiducial cross section, and which uses the minimum invariant mass of the possible pairs of $b$-jet and lepton as a proxy to the $\ttbar/tW$ interference effects.
These manifest at $m_{b\ell}^\mathrm{min}>150~\GeV$, where the kinematic threshold can be crossed by $tW$-like topologies where the $W$ boson not associated to the top quark resonance decays leptonically, and thus the lepton--$b$-jet system is not subject to the limit $m_{b\ell}<\sqrt{m_t^2-m_W^2}\sim 150~\GeV$.

It is precisely in the tail of the $m_{b\ell}^\mathrm{min}$ distribution that discrepancies are observed between the data and the various MC simulations tested, most notably between the DR and DS schemes for $tW$ with \POWHEG \hvq + \Pythia[8] as baseline for \ttbar.
Unfolded results are additionally presented for a ``search-like'' region at high transverse momenta and missing energy (and therefore also high $m_{b\ell}$), and for a ``hadronic $W$'' region that tightens the selection on the hadronic $W$ boson in order to significantly reduce the semi-leptonic $W+$jets background and its associated uncertainties.

The precise measurement of the leptonic differential cross sections performed by the ATLAS Collaboration in the $e\mu$ channel and discussed above~\cite{ATLAS:2025iza} also includes results unfolded to the fiducial volume corresponding to the $e\mu+b\bar{b}$ final state, i.e. without the requirement of any top quark or $W$ boson resonance.
This enables a direct test of the \POWHEG \bbfourl generator: it is found to perform very well, leading to substantially improved $\chi^2/$NDF results across all 1D and 2D observables, at a comparable level of agreement with the data to that of \MiNNLO.
In particular, the $\pT^{e\mu}$, $m^{e\mu}$ and $\Delta\phi^{e\mu}$ (although, interestingly, not $\Delta\phi^{e\mu}\times m^{e\mu}$) are slightly better modelled by \bbfourl than by \MiNNLO.
It is important to note that both \POWHEG programmes serve a different role, \MiNNLO producing more accurate \ttbar production matrix elements, while \bbfourl provides the complete NLO QCD description of the decays.
The observation that they both achieve improved descriptions of the data compared to \POWHEG \hvq is very promising, and further motivates the need for NNLO QCD corrections to \bbfourl.

\section{Fundamental physics at the threshold}
\label{sec:thresh}

\subsection{Quasi-bound states of top quarks}

The ATLAS and CMS Collaborations have both recently reported the observation of an excess of events near the top pair production threshold~\cite{CMS:2025kzt,ATLAS:2026dbe}, consistent with the formation of quasi-bound states (``toponium'').
These are expected in the SM to manifest near $\sim 2m_t-2~\GeV$, with a Coulombic QCD potential between the two non-relativistic top quarks arising from the exchange of soft gluons.
Contrary to lighter quarkonia, these quasi-bound states break due to the spontaneous electroweak decay of one of the constituent top quarks long before they have a chance to annihilate into one another.
Still, despite the very short lifetime of the toponium, an imprint of its QCD effects can be found in the decay products of the top quarks: a local enhancement of the cross section below the threshold, together with pseudo-scalar correlations of the top quark spin polarisations.
Both effects arise due to the dominant spin-singlet, colour-singlet $gg\to {}^1S_0^{[1]}\ttbar$ process.

The description of such toponium formation is formally absent from any MC simulation based on perturbative QCD (pQCD) and currently available to the ATLAS and CMS Collaborations, as it requires the separate framework of non-relativistic QCD (NRQCD) to make relevant predictions.
The toponium signal is therefore modelled separately, and added on top of the pQCD predictions of \POWHEG \hvq or \bbfourl.
In the CMS analysis~\cite{CMS:2025kzt}, the production of a pseudo-scalar resonance $\eta_t$ is modelled at leading order in \MADGRAPH as $gg\to\eta_t\to W^+bW^-\bar{b}$, with its mass, width and couplings tuned to reproduce NRQCD predictions of the \mttbar spectrum near threshold.
The ATLAS analysis~\cite{ATLAS:2026dbe}, on the other hand, starts from a leading order SM description of the $gg\to W^+bW^-\bar{b}$ matrix element, projects it onto a spin-singlet, and reweights each event according to the ratio of the NRQCD-inspired Green's function for the quasi-bound state over the free Green's function.
The Green's functions are computed in two dimensions, the energy of the toponium system and the characteristic momentum of the top quark.
The resulting model is referred to as ``Green's function-reweighted'', $\ttbar_\mathrm{GFRW}$.

Both experimental analyses proceed to a profile-likelihood fit of the pQCD \ttbar background and toponium signal to detector-level distributions of selected dileptonic events.
Two spin-sensitive observables are employed that bring crucial discrimination power between scalar and pseudo-scalar configurations of the top quarks, and therefore between the toponium signal and the pQCD \ttbar background.
The \mttbar distribution is fitted in slices of each spin-sensitive observable; in particular, the ATLAS analysis~\cite{ATLAS:2026dbe} limits it to $\mttbar < 500~\GeV$, avoiding spurious pulls and constraints on the pQCD \ttbar model from high-mass regions.
Both Collaborations report a significant excess over the pQCD \ttbar and background predictions, with a statistical significance beyond $5\sigma$, localised near threshold and behaving like spin-singlet toponium:
\begin{align}
  \sigma(\ttbar_\mathrm{GFRW}) &= 9.3\,^{+1.1}_{-1.0}\text{ (stat.)}\pm 0.8\text{ (syst.)}~\text{(pb)} \\
  \sigma(\eta_t) &= 8.8 \pm 0.5\text{ (stat.)}^{+1.1}_{-1.3}\text{ (syst.)}~\text{(pb)}
\end{align}
The measured inclusive cross sections are compatible with each other, but slightly higher than an NRQCD-inspired prediction of $6.43$~pb.
It is worth noting that refined theoretical predictions of these toponium effects are currently being developed, including a complete set of uncertainties on the prediction.
Both experimental results are limited by modelling uncertainties~\footnote{The ATLAS result~\cite{ATLAS:2026dbe} employs a different prescription for the breakdown of the total uncertainty returned by the profile-likelihood fit into its statistical and systematic components. This prescription is one that is compatible with results obtained from toy variations. When switching to the same uncertainty breakdown prescription as used in the CMS result~\cite{CMS:2025kzt}, comparable relative magnitudes of the statistical and systematic uncertainties are found. See the ATLAS paper~\cite{ATLAS:2026dbe} for more details.} on the pQCD \ttbar background and on the signal.
Remarkably, and in the absence of a tailored strategy for threshold or spin-sensitive effects, the ATLAS measurement of leptonic differential cross sections~\cite{ATLAS:2025iza} also exhibits some evidence (at the $3\sigma$ level) for toponium formation with a similar cross section (see the Appendix of that paper).

\subsection{Indirect measurement of the top quark Yukawa coupling}

Near the threshold, another class of interesting but subtle effects can be probed: the exchange of virtual Higgs bosons between the two top quarks.
Such interactions provide sensitivity to the square of the top quark Yukawa, $g_t^2$, and also appear in the distribution of \mttbar.
This provides an indirect way of extracting $g_t$, complementary to the $t\bar{t}H$ and $tH$ inclusive cross sections, top loops in Higgs production and decay to di-photons, or from off-shell Higgs bosons in the very rare $\ttbar\ttbar$ process.
A recent ATLAS measurement~\cite{ATLAS:2025ciy} performs a measurement of the top quark Yukawa modifier, defined as $Y_t=g_t/g_t^\mathrm{SM}$, using a semi-leptonic selection of \ttbar events.

The NLO EW effects are parameterised as a function of $(\mttbar, \cos\theta^*, Y_t^2)$, where $\cos\theta^*$ is the top quark scattering angle, and computed with the \HATHOR programme.
A template fit of the \mttbar distribution is performed at detector-level in terms of $Y_t^2$ rather than $Y_t$, to benefit from the linear dependence on $Y_t^2$.
Samples with negative $Y_t^2$ are added in the parameterisation to stabilise the fit.
Toponium effects are included using the LO-accurate simulations described above, and a conservative set of normalisation and shape uncertainties.
A value of
\begin{equation}
  Y_t^2=1.3 \pm 1.7
\end{equation}
is extracted from the data, corresponding to an upper limit in the $\mathrm{CL_s}$ construction at the $95\%$ confidence level of $Y_t<2.1$.
The results are therefore in agreement with the SM prediction for the top quark Yukawa, albeit with large uncertainties coming from the modelling of the \ttbar process and the jet energy scale.
The upper limit obtained is of comparable precision to that from the indirect measurement from $\ttbar\ttbar$.

\section{Conclusions}

The large datasets collected by the ATLAS and CMS Collaborations during Run~2 of the LHC have enabled very precise measurements of the \ttbar cross sections.
These have been obtained in different ways, including in dedicated low-pileup runs~\cite{CMS:2024ghc}, and in pure and well-understood dileptonic event selections~\cite{ATLAS:2025iza}.
Beyond doubly-resonant top quark production, the full decay chain $\ttbar\to W^\pm bW^\mp\bar{b}\to e^\pm\mu^\mp+b\bar{b}$ has become accessible~\cite{ATLAS:2025fhs}.
These off-shell topologies come with their own challenges, but these are being addressed with state-of-the-art modelling and simulations~\cite{ATL-PHYS-PUB-2025-036}.
Recent successes of the fundamental physics programme in the top quark sector at the LHC include an indirect measurement of the top quark Yukawa coupling~\cite{ATLAS:2025ciy} and the first observation of quasi-bound state (``toponium'') formation~\cite{CMS:2025kzt,ATLAS:2026dbe}.

These experimental results cement the role of the LHC as a precision machine, and we look forward to the exciting developments that will be brought about by the detailed study of the larger Run~3 dataset.

\section*{References}
\bibliography{BaptisteRavina}

\end{document}

